\documentclass[12pt]{iopart}

\usepackage{graphicx,dcolumn,bm,color,latexsym,amssymb}
\usepackage[latin1]{inputenc}
\usepackage{epstopdf}
\usepackage{subfigure}
\definecolor{Blue}{rgb}{0.0,0.0,1}
\definecolor{Red}{rgb}{1,0.0,0.0}
\definecolor{Green}{rgb}{0,0.5,0.0}

\begin{document}

\title{Scattering quantum circuit to measure Bell's time inequality
violation: a NMR demonstration using maximally mixed states}

\author{A. M. Souza$^{1,2}$, I. S. Oliveira$^2$ and R. S. Sarthour$^2$}
\address{$^1$Fakult\"at Physik, Technische Universit\"at Dortmund,
D-44221 Dortmund, Germany.}
\address{$^2$Centro Brasileiro de Pesquisas F\'{\i}sicas, Rua
Dr.Xavier Sigaud 150, Rio de Janeiro 22290-180, RJ, Brazil}
\ead{amsouza@cbpf.br} 

\begin{abstract}
\noindent  In 1985 A.J. Leggett and A. Garg proposed a Bell-like inequality 
to test the (in)compatibility between quantum mechanics the two fundamental concepsts. The first  
concept is the  ``macroscopic realism", that is, the quality of 
a physical property of a quantum system to be independent of observation at 
the macroscopic level, and the second concept is the ``noninvasive measurabilitly", that is, the 
possibility of performing a measurement without disturbing the subsequent evolution 
of a system. One of the key requirement for testing the violation of 
the Leggett-Garg 
inequality, or time Bell's inequality, is the ability to perform non-invasive measurements over a 
qubit state. In this paper we present a quantum scattering circuit which implements such a
measurement for maximally mixed states. The operation of the circuit is
demonstrated using liquid-state NMR in Chloroform, in which the time correlations of
one-qubit is measured on a probe (ancillary) qubit state. The results show
clearly a violation region, and are in excellent agreement with the quantum
mechanical predictions.
\end{abstract}

\pacs{ }
\maketitle


It has been recently reported by Palacios-Laloy and co-workers \cite{r1} an
experimental demonstration of Bell's time inequality violation in
superconductors, which is the first experimental test of the original proposal made
by Leggett and Garg in 1985 \cite{r2}. Experiments of this kind are
extremely important to test the foundations of quantum mechanics. The fundamental ideas 
addressed in the Leggett-Garg proposal were the ``macroscopic realism" and ``noninvasive measurability 
at the macroscopic level".  The 
former is the assumption that physical properties are independent of observation at 
macroscopic level. The later condition imples in the existence of measurements which can 
determine the state of a macroscopic system introducing an arbitrarily small perturbation that 
does not affect the subsequent dynamics. Studying these 
ideas on the superconducting current in a SQUID (Superconducting Quantum Inference Device) the 
authors conclude that quantum mechanics is incompatible with these two assumptions.  Besides 
that, Leggett-Garg inequality has also been related to the quantuness of 
computational algorithms \cite{mirikoshi} 
and the outcome of operators weakly measured \cite{wk1,wk2,wk3}, the 
so called weak values \cite{wm}. In  particular, the reference \cite{wk1} has shown that 
strange weak values, those values which exceed 
the range of eigenvalues associated to the observable in question, can only be found if and 
only if a Leggett-Garg inequality is violated.

On its usual form, Bell's inequality is 
violated by
multipartite quantum systems sharing an entangled state. Given a pure
entangled state, it is always possible to find a kind of inequality based on
local realism which is violated by quantum mechanics. However, it is
important to notice that some entangled mixed states do not violate standard
Bell's inequalities, unless additional local actions with classical
communication and post-selection (rejecting part of the original ensemble)
are performed \cite{genovese}. On another hand, a recent work \cite{ralph},
has suggested that quantum entanglement can be produced between thermal states
with nearly maximum Bell-inequality violation, even when the temperatures of
both systems approach infinity.

On the practical side, entanglement has been recognized as a key ingredient
for quantum computation and quantum communication \cite{r3}, and violation
of Bell's inequality can be used as a criterion of  efficiency for
quantum communication protocols \cite{r4a,r4b}. Moreover, whereas
multipartite quantum systems can exhibit non-local correlations introduced
by entanglement, one single quantum object, like a spin-1/2 particle, can
exhibit non-classical \emph{time-correlations} such as those addressed in 
the Leggett-Garg paper. In order to observe
such quantum time correlations during the evolution of a qubit and test the inequality, it is
necessary to perform the a noninvasive measurements. This
restriction imposes serious experimental difficulties.

The idea of Leggett-Garg is to measure time correlation functions of an observable $\mathcal{O}$
(which has eigenvalues $\pm 1$),  $C_{k,m}=\langle \mathcal{O}(t_{k})%
\mathcal{O}(t_{m})\rangle $, at different instants of time, $t_{k}$ and $%
t_{m}$, during which the system evolves under the action of a
time-independent Hamiltonian, $\mathcal{H}$, according to the Schroedinger
equation $\left\vert \psi \left( t\right) \right\rangle =\exp \left( -i%
\mathcal{H}t/\hbar \right) \left\vert \psi \left( 0\right) \right\rangle $ ($%
\left\vert \psi \left( 0\right) \right\rangle =\left\vert \psi \left(
t=0\right) \right\rangle =\left\vert \psi _{0}\right\rangle $).

The Leggett-Garg inequality, of the Wigner type \cite{MacroRealism}, states that for some chosen set of three
different instants of time, $t_{1}<t_{2}<t_{3}$, macroscopic realism imposes
that:

\begin{equation}
K\equiv C_{1,2}+C_{2,3}-C_{1,3}\leq 1  \label{eq0}
\end{equation}%
The condition for observing violations of this inequality, as demonstrated by 
Kofler and Brukner \cite{MacroRealism}, is that the initial
state of the system, $\left\vert \psi _{0}\right\rangle $, cannot be an
eigenvector of $\mathcal{H}$. The dichotomic observable $\mathcal{O}$ is
then defined in terms of the initial state as $\mathcal{O}\equiv 2\left\vert \psi
_{0}\right\rangle \left\langle \psi _{0}\right\vert -\mathcal{I}$, being $%
\mathcal{I}$ the identity matrix. Under time evolution, a measurement of
this observable, $M(\mathcal{O})$, will indicate whether the system is still
in $|\psi_0\rangle$ ($M(\mathcal{O})=+1$), or not ($M(\mathcal{O})=-1$).
Performing measurements for three different times such as $%
t_{2}-t_{1}=t_{3}-t_{2}=\Delta t$, the inequality (\ref{eq0}) takes the form 
\cite{MacroRealism}:

\begin{equation}
K=2\cos \left( \frac{\Delta E\Delta t}{\hslash }\right) -\cos \left( 2\frac{%
\Delta E\Delta t}{\hslash }\right) \leq 1  \label{eq4}
\end{equation}%
where $\Delta E$ is the energy separation between the qubit eigenvalues.
This inequality is clearly violated for $0<\Delta E\Delta t/\hbar <\pi /2$,
and is maximally violated for $\Delta E\Delta t/\hbar  = \pi /3$.

A great deal of effort has been put, since the work of Leggett-Garg \cite{r2},
to find ways to implement non-invasive measurements \cite{r5,r6,r7,r7a} and
verify the violation of the inequality (\ref{eq0}). In particular, the
proposal of \emph{weak measurements} of \cite{r6} has been implemented on
the experiment of Palacios-Laloy \cite{r1}. On spite of the successful
demonstration of Bell's time inequality violation in \cite{r1}, the
literature still lacks an example of a full quantum protocol with
noninvasive measurements. The proposal of this paper is to present an
scattering quantum circuit based on \cite{MacroRealism} which performs such
a test for a maximally mixed state, and to demonstrate its implementation using the well established NMR
quantum information processing experimental techniques \cite{r8,r9}.

The circuit proposed in this paper correlates a single microscopic qubit
(ancillary), to a quantum ensemble at infinite temperature. This is
similar to the quantum Schr\"odinger's cat paradox, where  the
macroscopic ``cat'' corresponds to a thermal state at infinite temperature, a state that 
is often believed to not exhibit quantum properties. The ancillary
qubit probes the time correlations of the ensemble without disturbing 
its subsequent dynamics.

An early example of NMR application to a quantum scattering circuit appeared
on a paper by Miquel et al., \cite{r10} in the context of measuring Wigner
functions on phase-space using NMR on liquid Trichloroethylene, a well known
3-qubit system. On a scattering circuit, a probe qubit (ancillary), prepared
in a known initial state, interacts with the system in such a way that a
measurement over its state after the interaction brings information about
the system state. For this, it is necessary that: (i) the input state of the
probing qubit to be known, and (ii) that the interaction can be controlled.
Figure \ref{fig1} shows a scattering circuit to obtain time correlation function of a qubit
by measuring the state of the probing qubit, which interacts with it. For a
non-invasive measurement, it is necessary to prepare the system input state
in such a way that the controlled interaction does not affect it. Such a
state can be worked out as follows: consider the input state,

\begin{equation}
\rho _{in}=\rho _{probe}\otimes \rho _{sys}=|0\rangle \langle 0|\otimes
|\psi _{0}\rangle \left\langle \psi _{0}\right\vert  \label{eq1}
\end{equation}%
where the first qubit on the left is the probe, entering the circuit on $%
|0\rangle $, and the other one is the system, prepared in the state $|\psi_0 \rangle $.
Upon the transformations shown on Figure \ref{fig1}, 
the output of the scattering circuit will be \cite{r10}:

\begin{equation}
\rho _{out}=|\varphi \rangle \left\langle \varphi \right\vert   \label{eq2}
\end{equation}%
where $\left\vert \varphi \right\rangle =\left\vert 0\right\rangle \otimes
\left( \mathcal{I}+U\right) |\psi _{0}\rangle +\left\vert 1\right\rangle
\otimes \left( \mathcal{I}-U\right) |\psi _{0}\rangle $, being $U=e^{i%
\mathcal{H}t_{m}/\hslash }\mathcal{O}e^{-i\mathcal{H}t_{m}/\hslash }e^{i%
\mathcal{H}t_{k}/\hslash }\mathcal{O}e^{-i\mathcal{H}t_{k}/\hslash }$. The
real part of the expected value of the spin $z$-component for the probing
qubit is:$\langle \sigma _{z}\rangle =\mathrm{Tr}\left\{ \rho
_{sys}U\right\} $ and therefore:

\begin{equation}
\langle \sigma _{z}\rangle =\left\langle \psi _{0}\right\vert e^{i\mathcal{H}%
t_{m}/\hslash }\mathcal{O}e^{-i\mathcal{H}t_{m}/\hslash }e^{i\mathcal{H}%
t_{k}/\hslash }\mathcal{O}e^{-i\mathcal{H}t_{k}/\hslash }|\psi _{0}\rangle
\label{eq3}
\end{equation}
which implies that:

\begin{equation}
\langle \sigma _{z}\rangle =\langle \mathcal{O}(t_{m})\mathcal{O}%
(t_{k})\rangle  \label{(eq3m)}
\end{equation}

According to \cite{MacroRealism}, any combination of two sates which are
orthogonal to the eigenvectors of $\mathcal{H}$ will lead to the violation
of the inequality (\ref{eq0}). Thus, both states $|0\rangle$ and $|1\rangle$ will lead to a violation 
when choosing $\mathcal{H}=\hslash \omega \sigma _{x}$, being $\sigma _{x}$, one of the 
Pauli matrices. It is interesting to note that the theoretical prediction of the 
quantity $K$, defined in (\ref{eq0}), as a function 
of $\Delta t$ is identical for both states and is given by the Equation (\ref{eq4}). Therefore, any 
statistical mixture of those states, $p_0 |0\rangle \langle0| + p_1 |1\rangle \langle1|$, will 
also violate 
the same inequality. Although the violation of Leggett-Garg Inequality  does not depend 
on the degree of mixedness, most states cannot  be used in a experimental test using the 
scheme proposed here because the application of the circuit  
of Figure \ref{fig1} in a general state will not perform a noninvasive measurement. Only the 
completely mixed  state ($p_0 = p_1 = 1/2$) does not 
undergo any change during the application of the circuit. Hence, this 
particular state is ideal for a macroscopic realism test. The probe qubit,
however, is initialized on a pure state. This situation
corresponds to the model of computation known as deterministic quantum
computation with one quantum bit (DQC1) \cite{dqc1}.

The ability of Nuclear Magnetic Resonance to generate quantum states and
unitary transformations of quantum circuits, is well established in the
literature \cite{r9,r10}. To implement the circuit of Fig \ref{fig1} and
demonstrate our proposal, we used a well known two-qubit NMR quantum
processor: the Chloroform molecule, CHCl$_{3}$. A liquid sample of 99.99\%
Carbon labeled diluted in deuterated acetone was used. The experiment
was performed at room temperature, in a Varian 500 MHz  Shielded NMR
spectrometer. It is worth noticing that the entire protocol lasts about $10$ ms, whereas the 
characteristic decoherence time for this system is $T_2 \approx 3$ s for Hydrogen 
and $T_2 \approx 0.8$ s for Carbon. Therefore, decoherence can be neglected during the application of the protocol.

\begin{figure*}[tbp]
\begin{center}
\includegraphics[width=1.1\linewidth,
  keepaspectratio]{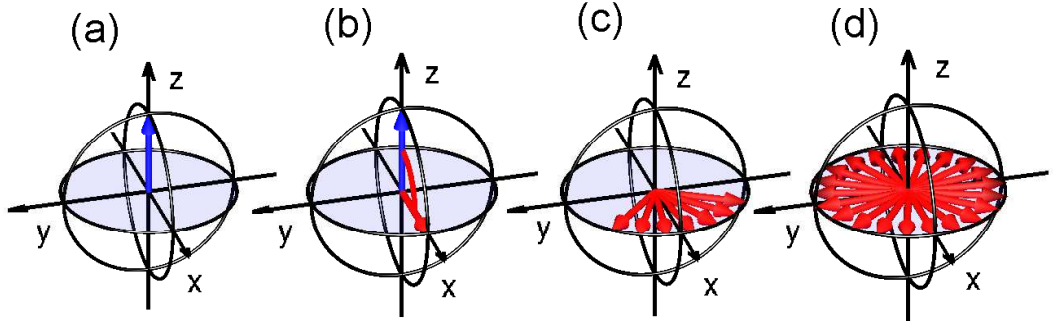}
\end{center}
\caption{ Evolution of the spins in the preparation of a maximally 
mixed state. (a) The Bloch vector of the spins are initially prepared in the state $|0\rangle$ (b) A $\pi/2$ 
rotation takes the Bloch vector from the state $|0\rangle$ to the $xy$ plane. (c) A field gradient in 
applied, since the fieled varies  along the $z$-axis, spins in different positions in the sample start 
to process with different angular velocities. (d) After some time the distribution of Bloch vectors are 
completely randomized, corresponding to a maximally mixed state.}
\label{ident}
\end{figure*}

\begin{figure*}[tbp]
\begin{center}
\includegraphics[width=0.5\linewidth,
  keepaspectratio]{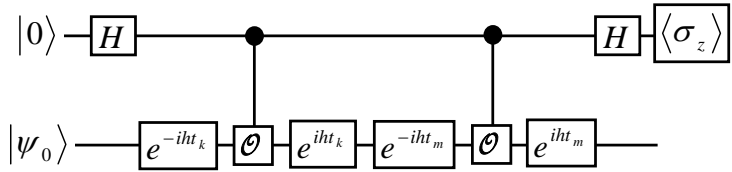}
\end{center}
\caption{ Quantum scattering circuit for measuring time correlation
functions $\langle \mathcal{O} (t_m) \mathcal{O} (t_k) \rangle$, at instants $t_m$ and $t_k$ where $h$ 
stands for $H/\hslash$. The correlation can be obtained by measuring the expected 
value $\langle\sigma_z\rangle$ of the ancillary
qubit.}
\label{fig1}
\end{figure*}

In order to observe time correlations and the violation of the
inequality (\ref{eq0}) we have used the Hamiltonian $\mathcal{H}=\hslash
\omega \sigma _{x}$. In NMR,
this Hamiltonian is simply implemented by radio-frequency pulses applied on
the resonance of the qubit spins. The tested ensemble corresponds to an ensemble of nuclear
spins 1/2 in a maximally mixed state, produced by a single $\pi/2$ pulse, followed by a field 
gradient, as illustrated in figure \ref{ident}. Due to the low polarization of the nuclear
spins at room temperature, the probe qubit is not initialized in a pure
state but rather in the pseudo-pure state $(1-\epsilon )\mathcal{I}%
/2+\epsilon |0\rangle \langle 0|$. Since $(1-\epsilon )\mathcal{I}/2$ is
not observed, the probe qubit in such a mixed state produces the same result
as it would be observed if the probe were in a pure state and the detection 
efficiency of the measurement apparatus were $\epsilon$. The data analysis here is   
analog to the post-selecting procedure, used in experiments with low efficiency detection \cite{genovese}.
In NMR,  post-selection is accomplished by normalizing the 
signal to a reference.

\begin{figure}[tbp]
\begin{center}
\includegraphics[width=0.65\linewidth,
  keepaspectratio]{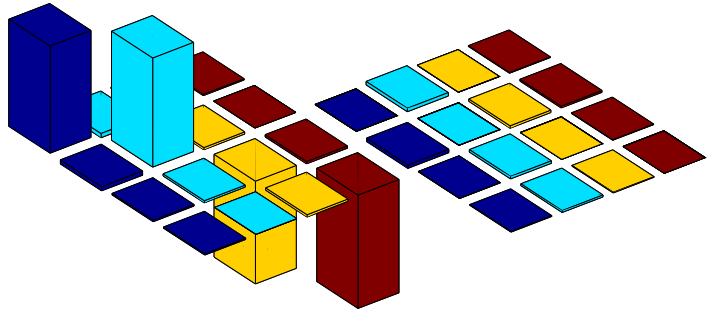}
\end{center}
\caption{Experimental tomographed input deviation density matrix  
$\Delta \rho =|0\rangle \langle 0|\otimes \mathcal{I}/2-\mathcal{I}/4$. The fidelity between the 
experimental and teoretical matrices is $\approx 0.99$.  }
\label{fig2}
\end{figure}

\begin{figure*}[tbp]
\begin{center}
\subfigure[ $ C_{12}(t)$ ] {\includegraphics[width=0.30 \linewidth,keepaspectratio]{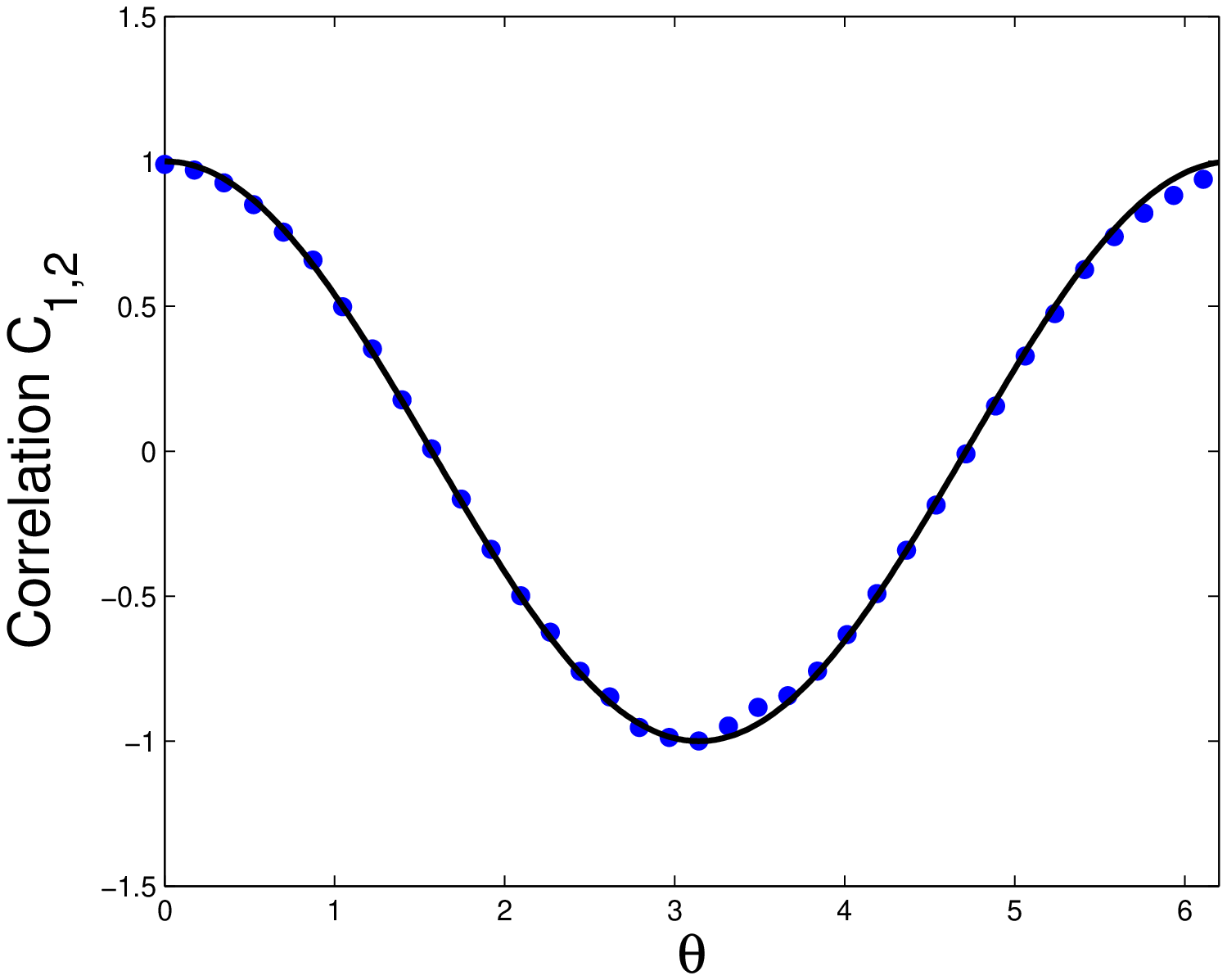}} 
\subfigure[ $ C_{23}(t)$ ] {\includegraphics[width=0.30\linewidth,keepaspectratio]{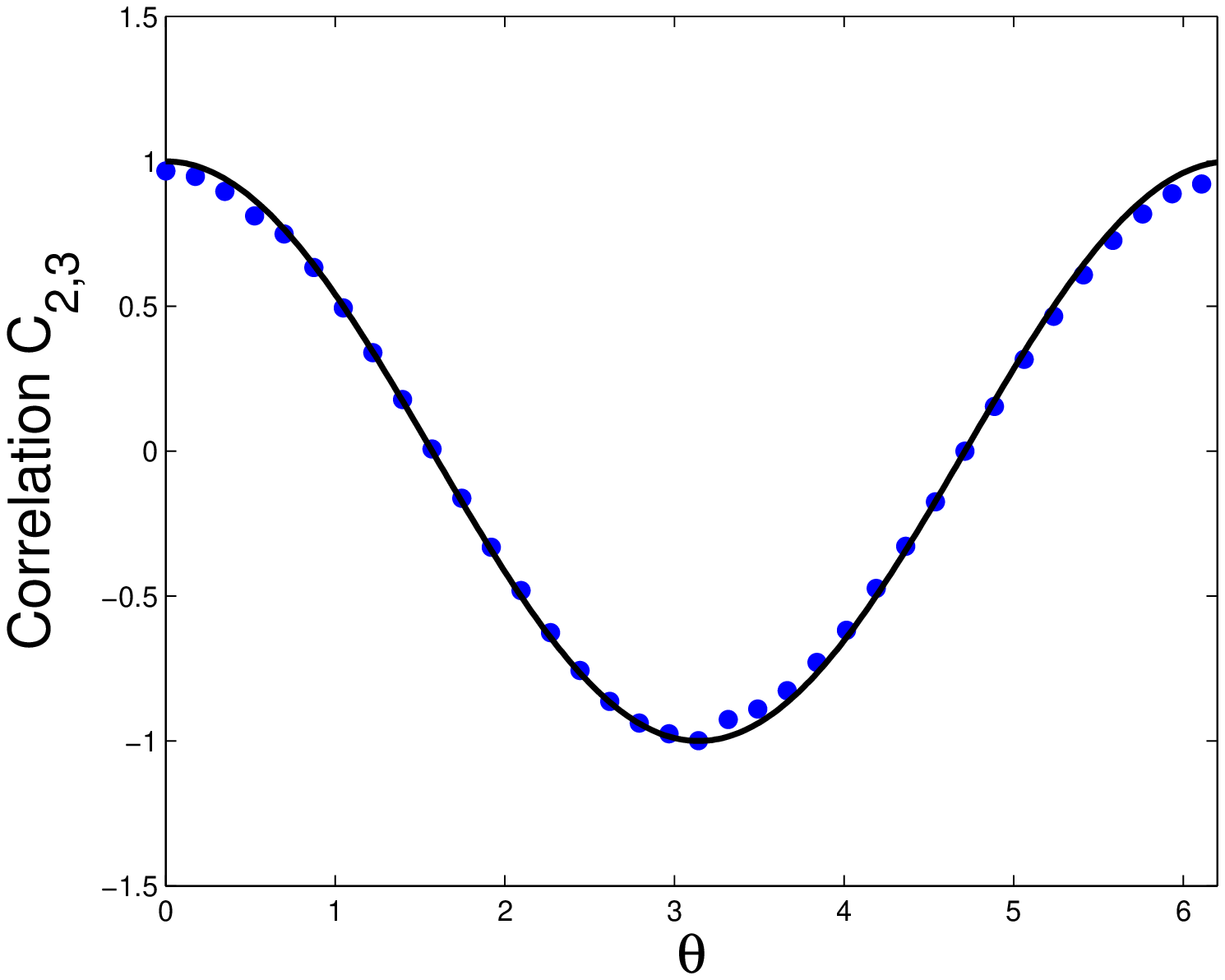}} 
\subfigure[ $ C_{13}(t)$ ] {\includegraphics[width=0.30 \linewidth,keepaspectratio]{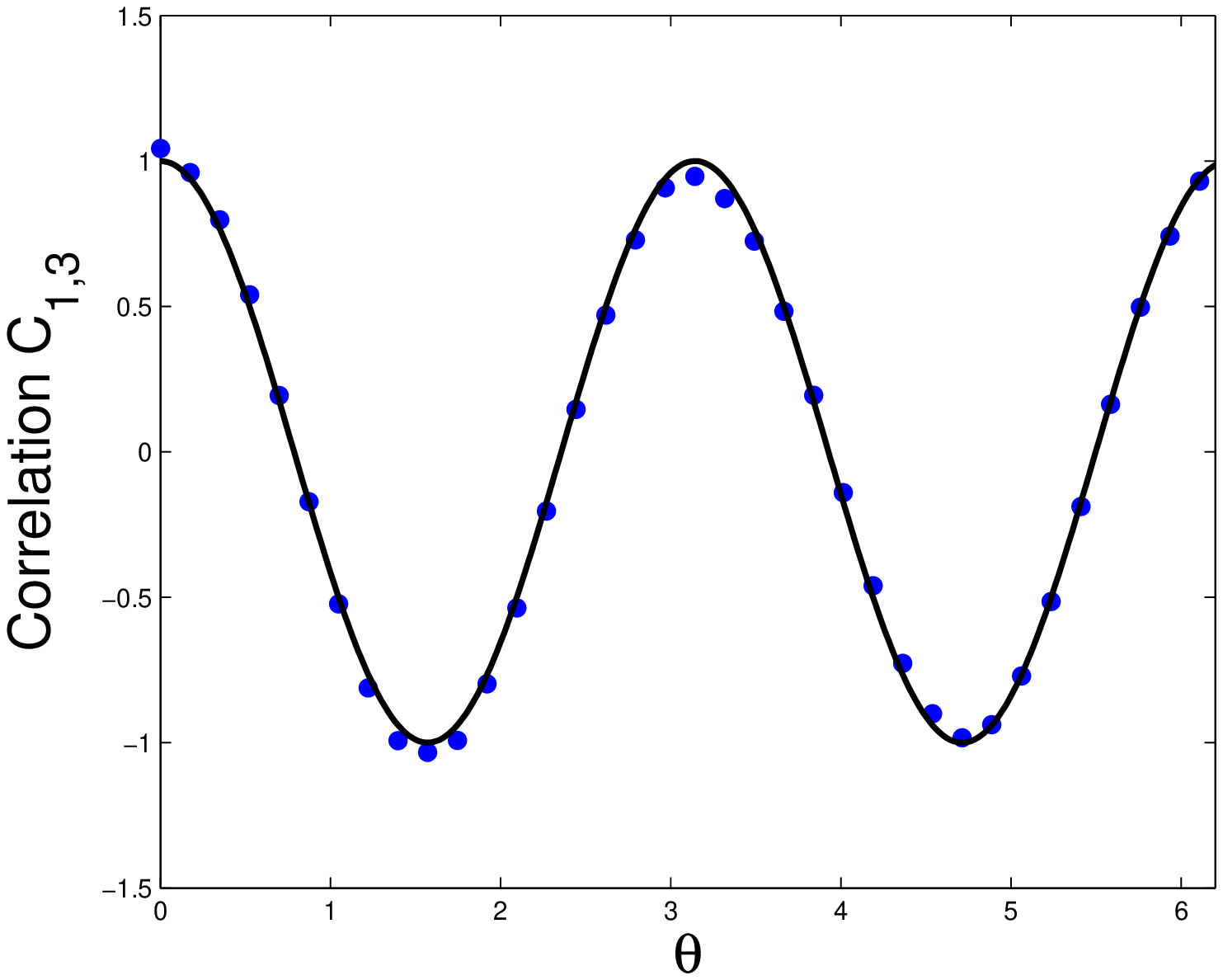}}\\
\end{center}
\caption{Correlation functions obtained in three different experiments. With extra
ancillary qubits it is possible to measure them all in a single run. The x-axis 
corresponds to a full $2\pi$ cycle;  $\theta$  stands  for $\Delta E \Delta t/\hslash$. }
\label{fig3}
\end{figure*}

\begin{figure}[tbp]
\begin{center}
\includegraphics[width=0.65\linewidth,
  keepaspectratio]{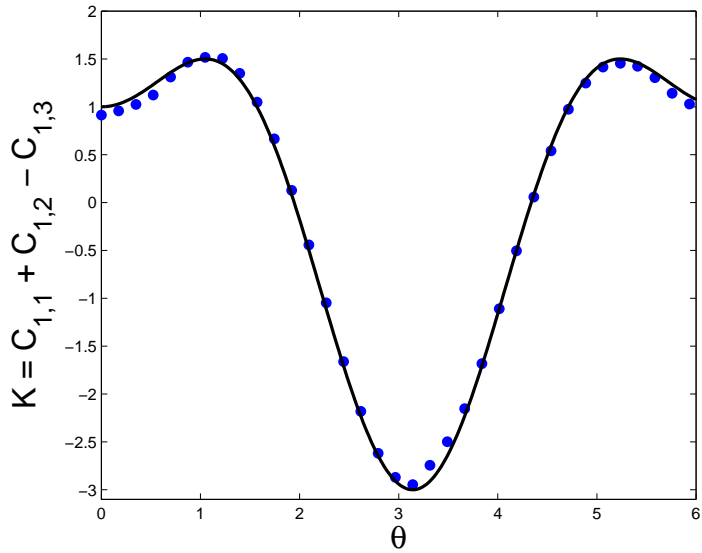}
\end{center}
\caption{Violation of Leggett-Garg inequality, Eq. (\protect\ref{eq0}). The x-axis  
corresponds to a full $2\pi$ cycle;  $\theta$ stands for $\Delta E \Delta t/\hslash$. The maximum 
violation occurs at $\pi/3$ and $5\pi/3$.}
\label{fig4}
\end{figure}

Figure \ref{fig2} shows the quantum state tomography of the system input
deviation density matrix $\Delta \rho =|0\rangle \langle 0|\otimes \mathcal{I%
}/2-\mathcal{I}/4$. Figure \ref{fig3} shows the correlation functions $C_{12}
$ (3a), $C_{23}$ (3b) and $C_{13}$ (3c) and the quantity $K$, defined in Eq. (\ref{eq0}), is shown 
in the Figure \ref{fig4}. The continuous line is the quantum
mechanical prediction. The results show
clearly a violation region, and are in excellent agreement with the quantum
mechanical predictions. However, the need of a normalization, or post-selection procedure,
introduces a detection loophole \cite{genovese}, and thus the violation of
realism can only the assumed upon the fair sampling hypothesis,
i.e. the hypothesis that sample of detected events is representative of the
entire system. Most of experiments testing standard Bell's inequalities are
run under this assumption. The remarkable fact about NMR is that the low spin
polarization implies the existence of a local and realistic hidden variable
model \cite{caves,nmrbell}. On the other hand, one must notice that truly 
quantum correlations can be present between the probe and the tested ensemble \cite{r13,nmrqc}.

NMR has been
extremely useful for testing fundamental ideas on quantum information. In
particular, studies of entanglement and fundamental quantum mechanics has been exploited in various 
NMR works (see for example \cite{nmrbell,nelson,laflamme}). Much less
exploited than multipartite entangled states are the time quantum
correlations proposed by Leggett and Garg. Understanding such correlations is
of great relevance for fundamental physics as well for quantum
computation and communication. From an experimental point of view, testing
time correlations is simpler, because it can be done over one single qubit.

In conclusion, the results shown on this paper reveals that even in the
extreme situation of a maximally mixed state, time quantum correlations
persists. We have proposed, and demonstrated with a NMR quantum
information processor, a scheme based on a microscopic qubit probe to test
time Bell's inequalities of quantum ensembles in a maximally mixed
state. Such a scheme could be useful to be implemented in recent experimental 
setups \cite{ralph,haroche2} where a thermal state can be correlated to pure 
microscopic qubits.  It is tempting to associate our results to the power of one qubit
computation \cite{dqc1,r12,r13}, and this is certainly a subject worth
pursuing in further experiments. Finally, we would like to mention that the
scattering quantum circuit presented here can be easily adapted to measure
the three correlation functions simultaneously, using more ancillary qubits.

\ack
We wish to acknowledge the support of CNPq, the Brazilian Quantum Information Project (INCT), the 
Rio de Janeiro Funding Agency (FAPERJ) and the PCI-CBPF program. AMS also acknowledges support of the 
German Funding Agency, DFG (Su 192/24-1).

\section*{References}
\bibliographystyle{unsrt}

\end{document}